\documentclass{article}
\pdfoutput=1 
\usepackage{spconf,amsmath,graphicx,bm,amssymb}
\usepackage{hyperref}
\usepackage{xcolor}
\usepackage[font=small,labelfont=bf]{caption}
\usepackage[square,sort&compress,comma,numbers]{natbib}
\usepackage{float}
\usepackage{dblfloatfix}

\title{ Self-Supervised Physics-Guided Deep Learning Reconstruction For High-Resolution 3D LGE CMR
}
     \name{\begin{tabular}{c}Burhaneddin Yaman$^{1,2}$, Chetan Shenoy$^{3}$,  Zilin Deng$^{1,2}$,  Steen Moeller$^{2}$, Hossam El-Rewaidy$^{4}$,\\ Reza Nezafat$^{4}$, and Mehmet Ak\c{c}akaya$^{1,2}$
 \end{tabular}
 }

 \address{$^{1}$ Electrical and Computer Engineering, University of Minnesota, Minneapolis, MN \\
      $^{2}$ Center for Magnetic Resonance Research, University of Minnesota, Minneapolis, MN\\
      $^{3}$ Medicine (Cardiology), University of Minnesota, Minneapolis, MN\\
      $^{4}$ Medicine (Cardiology), Beth Israel Deaconess Medical Center, Harvard Medical School, Boston, MA\\
            }

\begin{document}
%\ninept
\maketitle
\begin{abstract}
Late gadolinium enhancement (LGE) cardiac MRI (CMR) is the clinical standard for diagnosis of myocardial scar. 3D isotropic LGE CMR provides improved coverage and resolution compared to 2D imaging. However, image acceleration is required due to long scan times and contrast washout. Physics-guided deep learning (PG-DL) approaches have recently emerged as an improved accelerated MRI strategy. Training of PG-DL methods is typically performed in supervised manner requiring fully-sampled data as reference, which is challenging in 3D LGE CMR. Recently, a self-supervised learning approach was proposed to enable training PG-DL techniques without fully-sampled data. In this work, we extend this self-supervised learning approach to 3D imaging, while tackling challenges related to small training database sizes of 3D volumes. 
Results and a reader study on prospectively accelerated 3D LGE show that the proposed approach at 6-fold acceleration outperforms the clinically utilized compressed sensing approach at 3-fold acceleration.    
\end{abstract}

\begin{keywords}
Self-supervised learning, physics-guided deep learning, accelerated imaging, parallel imaging, cardiac MRI, late gadolinium enhancement
\end{keywords}

\section{Introduction}
\label{sec:intro}
Late gadolinium enhancement (LGE) cardiac MRI (CMR) is the clinical gold standard for identification of myocardial scar and fibrosis \cite{LGE_kim_circulation}. While 2D LGE CMR remains popular, 3D imaging offers improved SNR and spatial resolution \cite{LGE_3D_Saranathan,Akcakaya_LGE_Radiology}. Isotropic high-resolution 3D LGE further enables better delineation of small structures, potentially improving the assessment of left ventricular scar heterogeneity, right ventricular scar and left atrial scar \cite{Akcakaya_LGE_JMRI}. However, such high resolution 3D scans are prohibitively long, which is especially problematic in the presence of respiratory motion and contrast washout \cite{Akcakaya_LGE_Radiology}. Thus, image acceleration by means of parallel imaging \cite{Sense,Grappa} and compressed sensing (CS) \cite{lustig,Akcakaya2011,LGE_CS_Adluru_JMRI_StackofStars, LGE_CS_Adluru_JMRI,LGE_CS_Kamesh}, are necessary.

Recently, deep learning (DL) has been used for accelerated MRI due to its improved reconstruction quality over conventional approaches \cite{Hammernik,Hemant,JongChulYeeDLMagPhase,RAKI,Kamilov_Rare,LeslieYing_SPM}. Among such methods, physics-guided DL (PG-DL) techniques unroll conventional iterative algorithms consisting of data consistency (DC) and regularizer units for a fixed number of iteration \cite{Hammernik,Hemant}. The DC units utilize conventional linear methods, while the regularizer units are implicitly implemented using convolutional neural networks (CNNs) \cite{Hammernik,Hemant,Hosseini_JSTSP}. PG-DL approaches are typically trained in a supervised manner using fully-sampled data as ground-truth reference. However, acquisition of high-quality fully-sampled data is not possible in high-resolution 3D LGE CMR due to contrast washout \cite{Akcakaya_LGE_Radiology}. Thus, methods for training PG-DL reconstruction without fully-sampled data for improved LGE CMR is desirable. 

Recently, several methods have been proposed for this goal \cite{yaman_SSDU_MRM,JongChulyeCycleGan,Kamilov_Rare,LeslieYing_Unsupervised}. Among these, a recent approach named self-supervised learning via data undersampling (SSDU) \cite{Yaman_SSDU_ISBI,yaman_SSDU_MRM} trains neural networks without fully-sampled data by splitting available measurements into two disjoint sets. One of these is used for the DC units in the unrolled network and the other is used to define the loss function. SSDU was applied to 2D knee and 3D brain MRI \cite{yaman_SSDU_MRM}, showing matching quality to supervised training and improved quality over conventional methods. But in the latter setting, an inverse Fourier transform was applied along the fully-sampled frequency encoding direction, and the slices in this direction were processed individually through a 2D unrolled network. However, a truly 3D processing may further improve reconstruction quality since: 1) Using 3D CNNs in regularizer units may capture higher-dimensional correlations than the 2D case, 2) For self-supervised learning, 3D acquisitions provide a higher degree of freedom along three dimensions for selecting the two subsets for loss and DC units.

In this work, we sought to enable PG-DL reconstruction of 3D LGE CMR by extending the SSDU approach to 3D imaging. Results on prospectively 3-fold undersampled 3D isotropic high-resolution LGE CMR show that the proposed 3D self-supervised approach improves the reconstruction quality compared to the clinical CS approach \cite{Akcakaya_LGE_JMRI}, both at the acquisition acceleration rate, $R = 3$, and further retrospective acceleration $R = 6$. 

\section{Materials and Methods} 
\label{sec:meterialsandmethods}
\subsection{Unrolling Iterative Algorithms}
Let $\mathbf{y}_\Omega$ be the acquired k-space data with sub-sampling pattern $\Omega$, and $\mathbf{x}$ be the image to be recovered. The regularized least squares problem for MRI reconstruction is given as 
\begin{equation}\label{Eq:recons1}
\arg \min_{\bf x} \|\mathbf{y}_{\Omega}-\mathbf{E}_{\Omega}\mathbf{x}\|^2_2 + \cal{R}(\mathbf{x}),
\end{equation}
where  $\|\mathbf{y}_{\Omega}-\mathbf{E}_{\Omega}\mathbf{x}\|^2_2$ is a data consistency term, ${\bf E}_{\Omega}: {\mathbb C}^{M} \to {\mathbb C}^P$ is the multi-coil encoding operator containing coil sensitivities and partial Fourier matrix sampling, and  $\cal{R}(\mathbf{\cdot})$ is a regularizer. There exist several approaches to solve Eq. (\ref{Eq:recons1}), such as proximal gradient descent (PGD) or variable splitting with quadratic penalty (VSQP) \cite{fessler_SPM}. 

In VSQP, DC and regularizer units are decoupled using an auxilliary variable ${\bf z}$ that is constrained to be equal to ${\bf x}$. Then, Eq. (\ref{Eq:recons1}) is reformulated as
an unconstrained problem by imposing a quadratic penalty 
\begin{equation}\label{Eq:recons2}
\arg \min_{\bf x} \|\mathbf{y}_{\Omega}-\mathbf{E}_{\Omega}\mathbf{x}\|^2_2 + \mu \|\mathbf{x}-\mathbf{z}\|^2_2 + \cal{R}(\mathbf{x}),
\end{equation}
where $\mu$ is the penalty parameter. Eq. (\ref{Eq:recons2}) is solved via alternating minimization as 
\begin{subequations}
\begin{align}
& \mathbf{z}^{(i)} = \arg \min_{\bf z}\mu \lVert\mathbf{x}^{(i-1)}-\mathbf{z}\rVert_{2}^2 +\cal{R}(\mathbf{z})\label{Eq:recons3a}
\\
& \mathbf{x}^{(i)} = \arg \min_{\bf x}\|\mathbf{y}_{\Omega}-\mathbf{E}_{\Omega}\mathbf{x}\|^2_2 +\mu\lVert\mathbf{x}-\mathbf{z}^{(i)}\rVert_{2}^2\label{Eq:recons3b}
\end{align}
\end{subequations}
 where $\mathbf{z}^{(i)}$ is an intermediate variable and $\mathbf{x}^{(i)}$ is the desired image at iteration $i$. In PG-DL, these conventional iterative algorithms are unrolled for a fixed number of iterations, in which each iteration contains a DC and a regularizer unit. In PG-DL, regularizer sub-problem in Eq. (\ref{Eq:recons3a}) is solved  with neural networks and DC sub-problem in Eq. (\ref{Eq:recons3b}) is solved via conjugate gradient (CG) \cite{Hemant}. 
\begin{figure}[t]
  	\begin{minipage}[b]{1.0\linewidth}
  		\centering
  		\centerline{\includegraphics[width=8.5 cm]{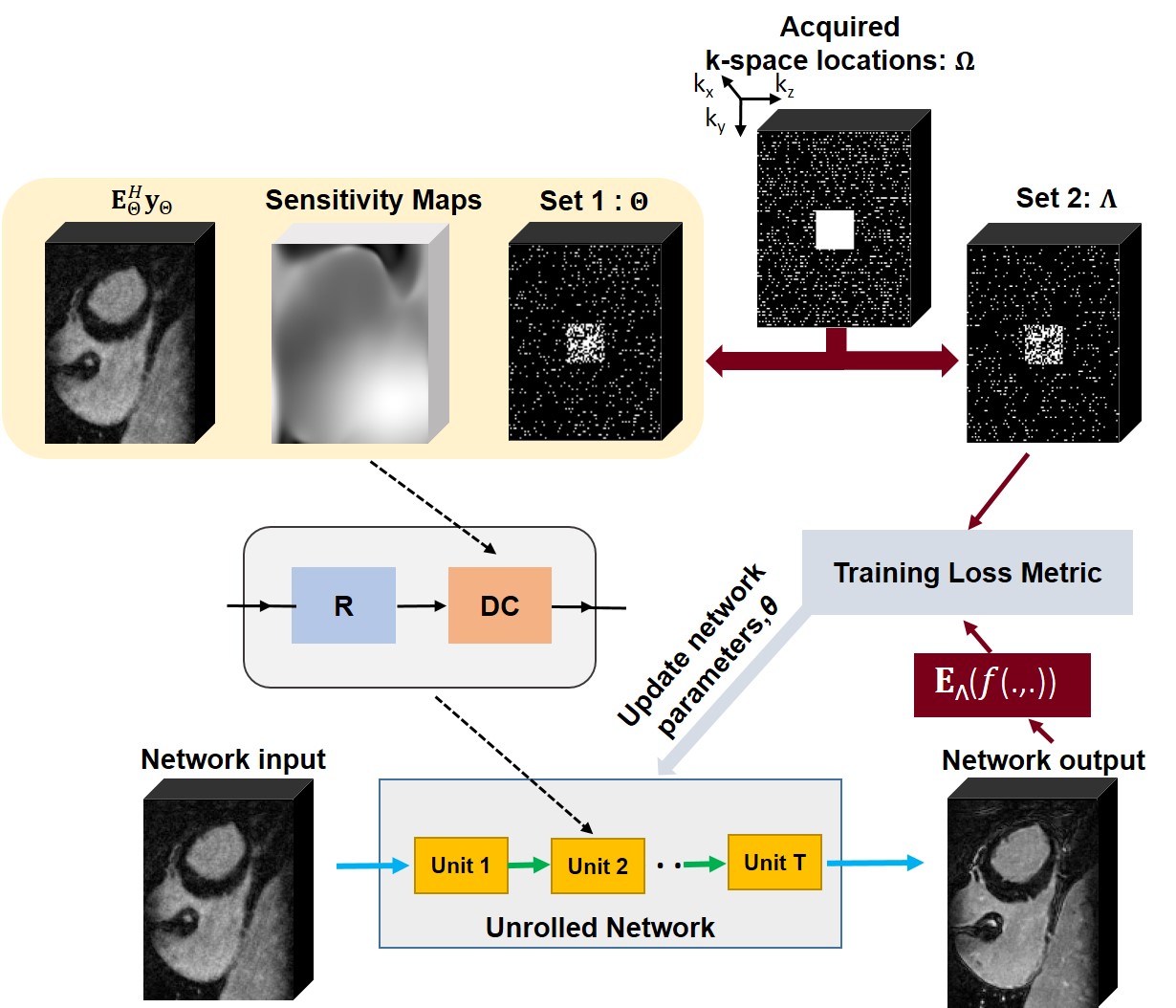}}
  	\end{minipage}

  	\caption{The self-supervised PG-DL training without fully-sampled data splits acquired sub-sampled k-space indices $\Omega$, into two disjoint sets, $\Theta$ and $\Lambda$. The first set of indices, $\Theta$, is used in the DC units of the unrolled network, while the latter set, $\Lambda$ is used to define the loss function for training. During training, the output of the network is transformed to k-space, and the available subset of measurements at $\Lambda$ are compared with the corresponding reconstructed k-space values. Based on this training loss, the network parameters are subsequently updated.}
  	\label{fig:3D_SSDU_Architecture}
  	\vspace{-.3cm}
\end{figure}

\begin{figure*}[b]

  	 \begin{center}
           \includegraphics[trim={0 0 0 0},clip, width=6.2 in]{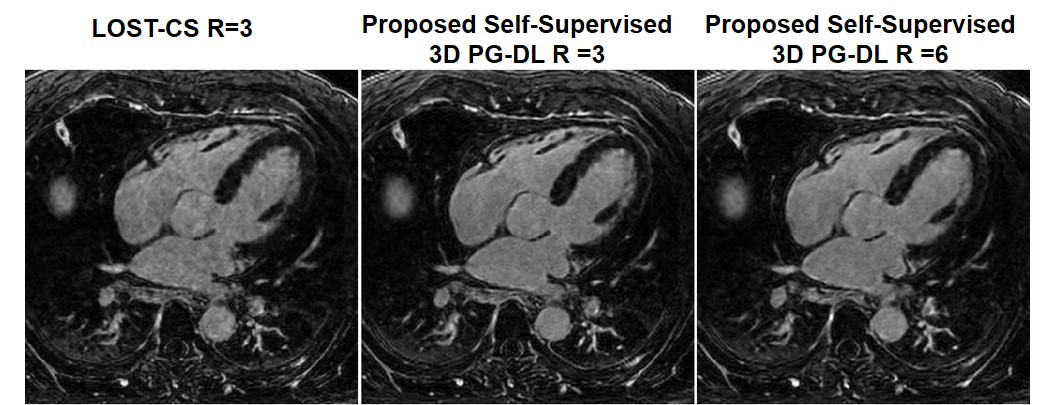}
     \end{center}
      \vspace{-.35cm}
  	\caption{Reconstruction results from a representative test slice without enhancement. LOST-CS was applied at the acquisition rate of 3, while the proposed 3D self-supervised PG-DL approach  was used at $R=3$ and 6. LOST-CS suffers from visible noise-like and incoherent residual artifacts. The proposed approach provides improved reconstruction at both $R = 3$ and 6. We further note that the proposed approach at $R=6$ only uses the data available at this rate for training, and does not have access to $R=3$ data. }
  	\label{fig:Cardiac_MRI_Fig1}
  	\vspace{-.3cm}
\end{figure*}

\begin{figure*}[t]
  	\begin{minipage}[b]{1.0\linewidth}
  		\centering
  		\centerline{\includegraphics[width=6 in]{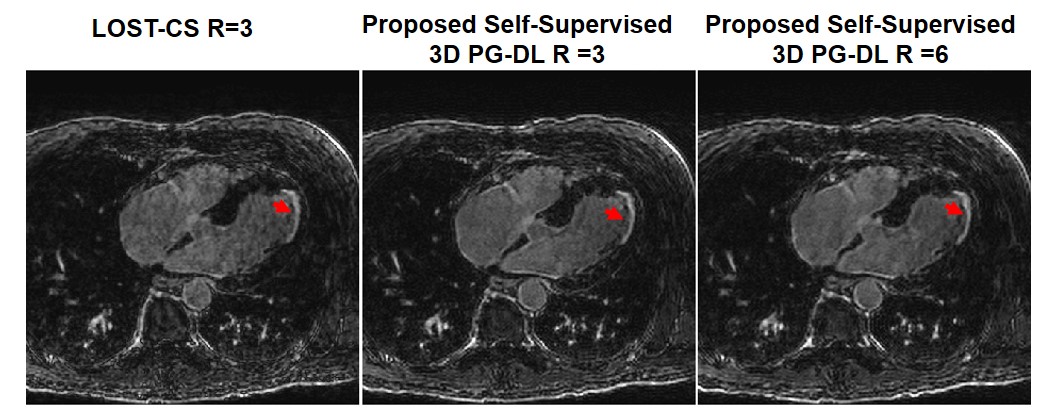}}
  			\vspace{-.2cm}
  	\end{minipage}
  	\caption{ 
  	Reconstruction results from a representative test slice with positive LGE. The proposed self-supervised PG-DL approach at both $R=3$ and 6 outperform LOST-CS reconstruction at $R=3$ by suppressing noise and residual artifacts. All reconstruction methods successfully identify LGE shown with red arrows.
  	}
  	\label{fig:Cardiac_MRI_Fig2}
  	\vspace{-.3cm}
\end{figure*}

\subsection{Supervised PG-DL Training}
Unrolled networks are trained end-to-end by minimizing a cost function between the network output and a reference. In supervised PG-DL, fully-sampled data is used as reference. Supervised PG-DL performs end-to-end training by minimizing an objective cost function given as
\begin{equation}
    \min_{\bm \theta} \frac1N \sum_{i=1}^{N} \mathcal{L}( {\bf y}_{\textrm{ref}}^i, \:{\bf E}_{\textrm{full}}^i f({\bf y}_{\Omega}^i, {\bf E}_{\Omega}^i; {\bm \theta})),
\end{equation}
where $N$ is the number of samples in the training database, $\mathcal{L}(\cdot, \cdot)$ is a loss function, ${\bf y}_{\textrm{ref}}^i$ is the fully-sampled k-space for subject $i$, $f({\bf y}_{\Omega}^i, {\bf E}_{\Omega}^i; {\bm \theta})$ is the output of the unrolled network for sub-sampled k-space data ${\bf y}_{\Omega}^i$ with the network being parameterized by ${\bm \theta}$, and ${\bf E}_{\textrm{full}}^i$ is the fully-sampled encoding operator that transforms network output to k-space. 

\subsection{Proposed 3D Self-Supervised PG-DL Training} \label{sec:23}
When the acquisition of fully-sampled data is either challenging or impossible, hindering supervised training of PG-DL approaches, SSDU enables training without fully-sampled data by splitting available undersampled measurements, $\Omega$ into two disjoint sets, $\Theta$ and $\Lambda$ to perform DC and defining loss. More formally, the cost function for training is modified to perform loss only on $\Lambda$ indices as
\begin{equation}
    \min_{\bm \theta} \frac1N \sum_{i=1}^{N} \mathcal{L}\Big({\bf y}_{\Lambda}^i, \: {\bf E}_{\Lambda}^i \big(f({\bf y}_{\Theta}^i, {\bf E}_{\Theta}^i; {\bm \theta}) \big) \Big).
    \label{eq:ssdu_loss}
\end{equation}

While the original SSDU was implemented with 2D networks \cite{yaman_SSDU_MRM}, 3D processing is desirable as discussed in Section \ref{sec:intro}. However, in addition to the difficulty of acquiring fully-sampled data for 3D scans, it is challenging to generate large databases of 3D acquisitions to train neural networks with a high number of parameters. We proposed to tackle these issues by extending SSDU to 3D processing, as shown in Figure \ref{fig:3D_SSDU_Architecture}. In addition to the selection of $\Theta$ and $\Lambda$ via Gaussian-weighted masking in three dimensions, we also tackle the issue of training data scarcity in databases by extracting multiple smaller 3D slabs from a whole heart acquisition of each subject. This is done by taking an inverse Fourier transform along the fully-sampled read-out direction direction, dividing the volume in this direction to smaller 3D sub-volumes, and processing the 3D k-space of these volumes.

End-to-end training was performed by unrolling iterative sub-problems in (\ref{Eq:recons3a})-(\ref{Eq:recons3b}) for $T=5$ iterations. DC units employed CG and regularizers used the same ResNet structure as in \cite{yaman_SSDU_MRM}, except 2D kernels of size $3 \times 3$ were replaced with 3D kernels of $3\times3 \times 3$.  Coil sensitivity maps were generated using ESPIRiT \cite{ESPIRIT}. Proposed 3D self-supervised PG-DL algorithm was trained using an Adam optimizer with a learning rate of $5\cdot 10^{-4}$ over 100 epochs by minimizing a normalized $\ell_1$ - $\ell_2$ loss \cite{yaman_SSDU_MRM}. All experiments for PG-DL approaches were performed using Tensorflow in Python.

\subsection{Imaging Experiments and Evaluation}
\label{ssec:invivodatasets} 
3D LGE CMR were acquired axially at 1.5T with a 32-channel coil array on 18 patients. The imaging protocols were approved by the local institutional review board, and written informed consent was obtained from all participants. Imaging parameters were: TR/TE = 5.2/2.6 ms, FOV = 320 $\times$ 320 $\times$ 100 $\textrm{ mm}^3$, resolution = 1.2 $\times$ 1.2 $\times$ 1.2 $\textrm{ mm}^3$, ACS = 40 $\times$ 24, prospective acceleration $R = 3$ with random $k_y - k_z$ sub-sampling \cite{Akcakaya_LGE_JMRI}.

The prospectively subsampled 3D LGE data was further retrospectively subsampled to $R=6$ by keeping a 24 $\times$ 24 ACS region in the $k_y-k_z$ plane using a random uniform undersampling pattern. Due to the small training database size, smaller 20 $\times$ 270 $\times$ 102 3D slabs for training were generated as described in Section \ref{sec:23}. Training was performed on 200 sub-volumes from 10 different subjects, and testing was performed on the whole volume of 8 different subjects.

The proposed 3D self-supervised PG-DL approach was performed at $R \in \{3,6\}$. For $R=6$ training, only data available at this rate were used. Results were compared with a clinically-used CS approach \cite{Akcakaya_LGE_JMRI}, LOST (Low-dimensional structure self-learning and thresholding) applied at $R=3$. We note that results could not be compared with supervised learning due to a lack of fully-sampled data. Quantitative metrics such as PSNR or SSIM were also not available due to the lack of ground-truth data. Qualitative assessment of the reconstruction image quality was evaluated by an experienced cardiologist using evaluation criteria of perceived SNR, blurring and overall image quality. The reader was blinded to the reconstruction methods and $R$. Evaluations were based on a 4-point ordinal scale for blurring (1: no blurring, 2: mild blurring, 3: moderate blurring, 4: severe blurring), perceived SNR (1:excellent, 2: good, 3: fair, 4: poor), and overall image quality (1: excellent, 2: good, 3: fair, 4: poor). Wilcoxon signed-rank test was used to evaluate the scores with a significance level of $P < 0.05$.

\section{Results}
\label{ssec:results} 
Figure \ref{fig:Cardiac_MRI_Fig1} shows reconstruction results on a representative test slice with negative LGE. The proposed 3D self-supervised approach was applied at both prospective acceleration $R=3$ and further retrospective acceleration $R=6$, while LOST-CS was applied only at prospective acceleration $R=3$. LOST-CS reconstruction shows a mixture of noise-like amplification and incoherent aliasing artifacts due to random undersampling, especially in the blood pool. The proposed 3D self-supervised approach at both $R = 3$ and 6 suppress these artifacts further and achieves improved reconstruction quality. 

Figure \ref{fig:Cardiac_MRI_Fig2} shows reconstruction results of LOST-CS and the proposed 3D self-supervised approach on a subject with positive LGE (marked by red arrows). Similar observations apply in this case, with the proposed method at $R = 3$ and 6 suppressing the residual artifacts in LOST-CS at $R = 3$, while also allowing a sharper depiction of the myocardium-blood border. All approaches show the enhancement region clearly.

\begin{figure}[!b]
  	\begin{minipage}[b]{1.0\linewidth}
  		\centering
  		\centerline{\includegraphics[width=8.5 cm]{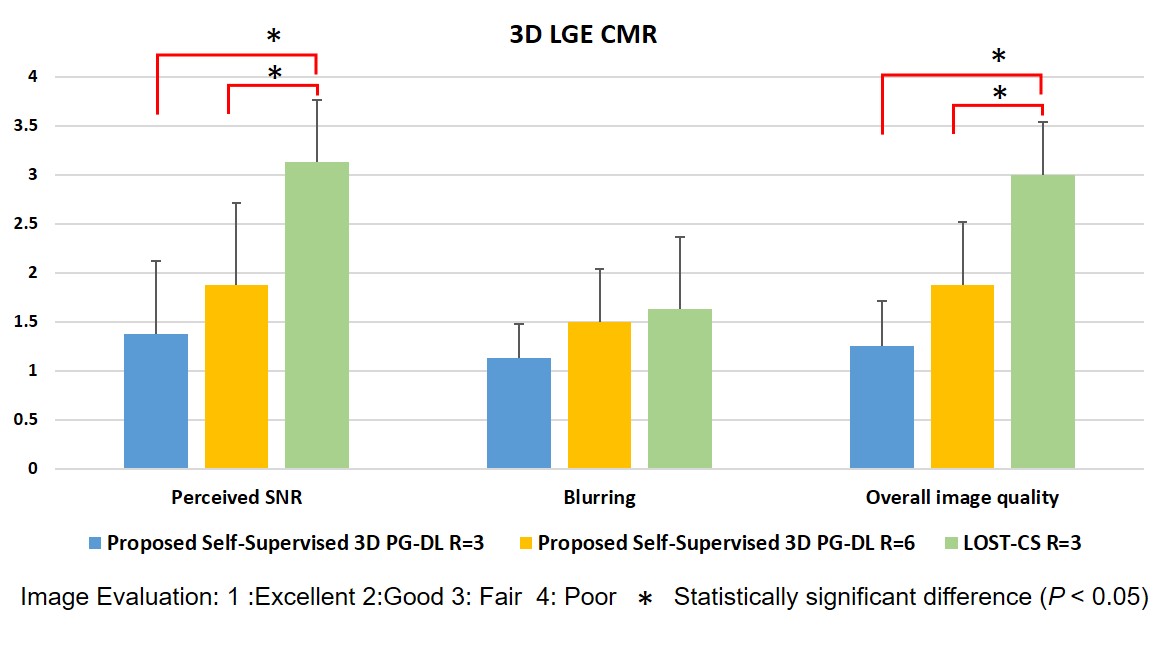}}
  		  	\vspace{-.2cm}
  	\end{minipage}
  	\caption{The image reading results from the clinical reader study for 3D LGE CMR. Evaluations were based on a 4-point ordinal scale (1:best, 4: worst). Bar-plots depict average and standard deviation across test subjects, with * showing statistically significant differences. For blurring, proposed self-supervised 3D PG-DL at $R=3$ and 6 were rated higher than LOST-CS at R=3, though the differences were not significant. For perceived SNR and overall image quality, proposed self-supervised 3D PG-DL at $R=3$ and $R=6$ were both rated statistically better than LOST-CS at $R=3$.}
  	\label{fig:Reader_Study}
  	\vspace{-.3cm}
\end{figure}

Figure \ref{fig:Reader_Study} summarizes the reader study for 3D LGE CMR dataset. Bar-plots show the average and standard deviation of reader scores across the test dataset. For blurring, all methods were statistically in good agreement, while the proposed 3D self-supervised at $R=3$ and $R=6$ was rated higher than LOST-CS at $R=3$. For both perceived SNR and overall image quality, proposed 3D self-supervised at $R=3$ and 6 were rated statistically better than LOST-CS at $R=3$ which is the current clinical approach. 
\section{Discussion and Conclusion}
\label{sec:discussion}
The proposed 3D self-supervised PG-DL reconstruction enables training neural networks without fully-sampled data for 3D volumes, by splitting available measurements into two disjoint sets, and using one of these in DC units and the other in defining loss. Moreover, we proposed to tackle the issue of having a small number of training subjects in the database for 3D applications by splitting the whole volume for each subject into sub-volumes and training over these sub-volumes. Results on 3D LGE CMR shows that the proposed approach significantly improves upon the state-of-the-art CS methods. 

Training of 3D PG-DL methods has gained interest recently due to their ability to capture higher-dimensional interactions. Several supervised PG-DL approaches have been proposed for 3D training \cite{Kellman_TCI_GpuMemory, CineNet}, either by using fully-sampled data or a surrogate reconstruction, such as CS or parallel imaging, as reference. However, the former is difficult in many CMR acquisitions due to scan length, contrast washout or motion; while the latter inherently limits the potential of PG-DL to existing reconstruction strategies. While self-supervised learning can tackle these issues by efficient utilization of acquired measurements, small training dataset size and GPU-memory issue of fitting large volumes \cite{Kellman_TCI_GpuMemory} are common challenges for both supervised and self-supervised PG-DL approaches. The proposed approach of splitting large volumes into sub-volumes provides an alternative solution to both of these issues.   
\section{acknowledgements}
\label{sec:acknowledgments}
This work was partially supported by NIH R01HL153146, P41EB027061, U01EB025144; NSF CAREER CCF-1651825.
% -------------------------------------------------------------------------
\section{Compliance with Ethical Standards }
The research study was performed in line with the principles of the Declaration of Helsinki. Approval was granted by the Institutional Review Board of the Beth Israel Deaconess Medical Center.
\bibliographystyle{IEEEbib}
\bibliography{reference}
\end{document}